\documentclass[twocolumn,showpacs,preprintnumbers]{revtex4}
\usepackage{graphicx}
\usepackage{dcolumn}
\usepackage{bm}
\usepackage{color}

\input{tcilatex}
\begin{document}

\title{Probing Half-odd Topological Number with Cold Atoms in a Non-Abelian
Optical Lattice}
\author{Feng Mei$^{1,3}$}
\author{ Shi-Liang Zhu$^{2}$}
\author{ Xun-Li Feng$^{1}$}
\author{Zhi-Ming Zhang$^{1}$}
\email{zmzhang@scnu.edu.cn}
\author{C. H. Oh$^{3}$}
\email{phyohch@nus.edu.sg}
\affiliation{$^{1}$Laboratory of Photonic Information Technology, LQIT $\&$ SIPSE, South
China Normal University, Guangzhou 510006, China\\
$^{2}$Laboratory of Quantum Information Technology and SPTE, South China
Normal University, Guangzhou, China\\
$^{3}$Centre for Quantum Technologies and Department of Physics, National
University of Singapore, 3 Science Drive 2, Singapore 117543, Singapore}
\date{\today }

\begin{abstract}
We propose an experimental scheme to probe the contribution of a single
Dirac cone to the Hall conductivity as half-odd topological number sequence.
In our scheme, the quantum anomalous Hall effect as in graphene is simulated
with cold atoms trapped in an optical lattice and subjected to a
laser-induced non-Abelian gauge field. By tuning the laser intensity to
change the gauge flux, the energies of the four Dirac points in the first
Brillouin zone are shifted with each other and the contribution of the
single Dirac cone to the total atomic Hall conductivity is manifested. We
also show such manifestation can be experimentally probed with atomic
density profile measurements.
\end{abstract}

\maketitle

\section{Introduction}

With easy production of graphene, this strictly two-dimensional material has
evoked strong interest in condense matter and high energy physics in the
past few years, for instance, the integer quantum Hall effect (QHE) has been
observed in graphene {\color{blue}\cite{graphene1}}. Remarkably, the
transverse Hall conductivity has an anomalous form: $\sigma
_{xy}=4(n+1/2)e^{2}/h$ ($n=0,\pm 1,...$), where $n$ is the Landau level (LL)
index and the factor 4 accounts for double valley and double spin
degeneracy. Note that the conductance sequence is shifted with respect to
the standard QHE sequence by 1/2. This unusual sequence is now well
understood as arising from the fact that the zero energy LL has half of the
higher LLs degeneracy {\color{blue}\cite{graphene2}}, and the Hall
conductivity for each Dirac fermion is half-integer quantized as $%
(n+1/2)e^{2}/h,$ ($n=0,\pm 1,...$) {\color{blue}\cite{HI}}. Due to the
presence of fermion doubling theorem {\color{blue}\cite{fd}}, which states
that for a time reversal invariant system Dirac points must come in pairs,
the contribution of the single Dirac cone to the total Hall conductivity of
the graphene is hidden. Even though one can find some lattice models which
have an odd number of massless Dirac cones to avoid the fermion doubling
theorem, the final Hall conductivity of \ the model is still integer
quantized because of the hidden massive Dirac fermions {\color{blue}\cite%
{hdf}}. In fact, the key reason lies in the famous
Thouless--Kohmoto--Nightingale--den Nijs (TKNN) formula which guarantees
that the total Hall conductivity is quantized as some integer numbers if the
Fermi energy stays in the energy gap {\color{blue}\cite{TKNN}}. So far
experiment is still lacking to demonstrate the half-integer Hall
conductivity contributed from a single Dirac cone.

Very recently, Watanabe and colleagues proposed a model to show that each
Dirac cone does indeed contribute to the Hall conductivity with half-odd
topological numbers {\color{blue}\cite{graphene3}}. In their work, they
constructed a lattice model in graphene with complex second nearest-neighbor
hopping integral among some particular lattice sites. The most interesting
feature of the model is that the energies of the two Dirac points are
shifted with each other, but their massless Dirac cones are still preserved.
Through computing the topological number for the shifted Dirac cones, the
half-integer contribution to the Hall conductivity from single Dirac cone
can be demonstrated. However, as the authors themselves pointed out in their
work, the lattice model is fictitious and it is extremely hard to realize
the unusual second nearest-neighbor hopping proposed there.

On the other hand, optical lattices populated with cold atoms offer a very
promising alternative avenue to explore rich fundamental phenomena of
condensed matter physics {\color{blue}\cite{QS}}. For example, ultracold
atoms in the optical lattices can be used to simulate the Bose-Hubbard model
and the fractional quantum Hall effect, etc. {\color{blue}\cite{Jaksch}}. In
analogy to the graphene, it is also proposed that the Dirac fermions can be
simulated and detected by the ultracold atoms in the honeycomb optical
lattice {\color{blue}\cite{Zhu2007}} and then demonstrated in details that
the massive and massless Dirac particles can be realized in realistic
conditions {\color{blue}\cite{Lee}}. In addition to the honeycomb lattices,
the Dirac fermions can also be generated from a line-centered-square lattice
or a lattice with $T_{3}$ symmetry {\color{blue}\cite{Shen}}. Furthermore,
the Dirac cones can be realized with a square lattice subjected to an
artificial non-Abelian gauge potential. The artificial gauge potentials
acting on neutral atoms have recently attracted considerable interest {%
\color{blue}\cite{dalibard}}. The gauge potential can lead to an effective
magnetic field and provide opportunities to simulate the physics of charges
subjected to magnetic fields. Many recent works have shown that the Abelian
and non-Abelian gauge fields can be realized using optical dressing {%
\color{blue}\cite{dressed} }or laser-assisted hopping {\color{blue}\cite{hop}
}in optical lattice for observing the Hofstadter butterfly {\color{blue}\cite%
{HB}}, atomic QHE {\color{blue}\cite{AHE1,AHE2,AHE3}}, topological
insulators {\color{blue}\cite{TI}}, axion electrodynamics {\color{blue}\cite%
{AE} }and Majorana fermion {\color{blue}\cite{MF}}. Both Abelian and
non-Abelian gauge fields in cold atoms have recently been generated
experimentally {\color{blue}\cite{NIST}}.

Motivated by these ongoing developments, we propose an experimental scheme
to manipulate single Dirac cone and probe its contribution to Hall
conductivity as half-odd topological numbers with cold atoms in the optical
lattices subjected to an artificial non-Abelian gauge potential. We devise a
distinct method to engineer the Dirac cone through changing the non-Abelian
gauge flux in optical lattice. In the $\pi $-flux regime, there will be four
zero energy Dirac points in the first Brillouin zone which leads to the
anomalous Hall effect as in graphene. The key point in our paper is that,
through changing the non-Abelian gauge flux, the energies of the four Dirac
points are shifted with each other, but their positions and massless Dirac
cone characters remain intact. By manipulating the Dirac cones and
calculating the corresponding topological Chern numbers, the contribution of
the single Dirac cone to the total atomic Hall conductivity as half-odd
topological numbers is demonstrate indirectly. In our model, different gauge
fluxes can be easily adjusted through tuning the laser field intensity.
Furthermore, we show that the atomic Hall conductivity could be detected
with the standard density profile measurement used in cold atomic systems.

\begin{figure}[tbp]
\includegraphics[width=8cm]{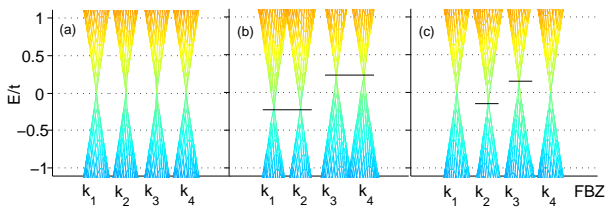}
\caption{(Color online) The dispersion relation near the four Dirac cones in
the FBZ for the different parameters $\protect\alpha $ and $\protect\beta $.
(a) $\protect\alpha =\protect\beta =\protect\pi /2$; (b) $\protect\alpha =%
\protect\pi /2+0.1$, $\protect\beta =\protect\pi /2$; (c) $\protect\alpha =%
\protect\pi /2+0.05$, $\protect\beta =\protect\pi /2-0.05$.}
\end{figure}

\section{SUMULATING AND ENGINEERING DIRAC CONE}

Let us consider a two-component ultracold fermion gas trapped in a two
dimensional square optical lattice with the lattice sites ($x=ma$, $y=na$),
where $a$ is the lattice spacing, $t$ is the nearest-neighbor hopping
strength and the integers $m$ and $n$ are the position of the lattice site.
We assume that the atomic collisions are rendered negligible through
adjusting Feshbach resonance {\color{blue}\cite{AS}}. The optical lattice
setup is subjected to a gauge potential as $\vec{A}=(\alpha \sigma _{y},$ $%
\beta \sigma _{x}+2\pi \Phi m)$, where $\sigma _{x(y)}$ is the Pauli
operator of the two atomic components labelled by two atomic sublevels of a
hyperfine manifold, $\Phi $ is the uniform Abelian magnetic flux per unit
cell and we set $c=\hbar =e=a=1$. This gauge potential may be generated
using a two dimensional optical superlattice based on laser-assisted
spin-dependent hopping {\color{blue}\cite{AHE2,AHE3,AE,NAB}}. The gauge
fluxes $\left( \alpha \text{,}\beta \right) $ are determined by the Rabi
frequencies of the two-photon off-resonant transition and can be tuned via
changing the laser field intensity {\color{blue}\cite{AHE2,AHE3,AE,NAB}}.
The tight-binding Hamiltonian of the system is written as 
\begin{widetext}
\begin{equation}
H=-t\sum\limits_{m,n}\left( c_{m+1,n}^{\dagger }e^{i\alpha \sigma
_{y}}c_{m,n}+c_{m,n+1}^{\dagger }e^{i\left( \beta \sigma _{x}+2\pi \Phi
m\right) }c_{m,n}\right) +H.c,
\end{equation}%
\end{widetext}where $c_{m,n}$ is a 2-component field operator defined on a
lattice site ($m,n$). We firstly study the system in the non-Abelian gauge
potential when the uniform Abelian gauge flux $\Phi =0.$ By diagonalizing
the above Hamiltonian, one can get the single-particle energy spectra as $%
E_{k}=h_{0}\left( k\right) \pm \sqrt{h_{x}^{2}\left( k\right)
+h_{y}^{2}\left( k\right) }$, where $h_{0}\left( k\right) =-2t(\cos (\alpha
)\cos (k_{x})+\cos (\beta )\cos (k_{y}))$, $h_{x}\left( k\right) =-2t\sin
(\beta )\sin (k_{y})$ and $h_{y}\left( k\right) =-2t\sin (\alpha )\sin
(k_{x})$. When the gauge fluxes $\alpha =\beta =\pi /2$ ($\pi $-flux
regime), this dispersion relation leads to four Dirac points around which
the dispersion relation is linear {\color{blue}\cite{AHE3}}. The four Dirac
points are located at $\mathbf{k}_{1}=(0,0)$, $\mathbf{k}_{2}=(0,\pi )$, $%
\mathbf{k}_{3}=(\pi ,0)$, $\mathbf{k}_{4}=(\pi ,\pi )$ in the first
Brillouin zone (FBZ) with zero energies. Note that this $\pi $-flux gauge
potential is in the Abelian regime because its Wilson loop $|W|=2$ {%
\color{blue}\cite{AHE2}}$.$ Interestingly, the relative energies of the four
Dirac points could be shifted through changing the gauge flux from the
Abelian to the non-Abelian regime ($|W|<2$). This can be seen clearly by
expanding the Hamiltonian in Eq.(1) around the four Dirac points. Apart the
conventional linear dispersion relation related to the Dirac Hamiltonian, we
find that the energies of Dirac points become 
\begin{eqnarray}
\delta E_{1(4)} &=&\pm 2t(\cos (\alpha )+\cos (\beta )),  \nonumber \\
\delta E_{2(3)} &=&\pm 2t(\cos (\alpha )-\cos (\beta )).
\end{eqnarray}%
In the case of Abelian $\pi $-flux regime ($\alpha =\beta =\pi /2$), the
energies of the four Dirac points are zero, which agrees with the previous
description. However, if the gauge fluxes have been changed into the
non-Abelian regime $\alpha =\pi /2+\triangle \alpha $ and $\beta =\pi
/2+\triangle \beta $, their energies are $\delta E_{1(4)}=\mp 2t(\sin
(\triangle \alpha )+\sin (\triangle \beta ))$ and $\delta E_{2(3)}=\mp
2t(\sin (\triangle \alpha )-\sin (\triangle \beta ))$ and would be shifted
with respect to each other. As shown in Fig. 1, the energies of the four
Dirac points vary with the gauge fluxes but the massless Dirac cone
character is still kept. In particular, we shift the energies of four and
two Dirac points respectively in Fig.1(b) and (c).

\section{HALF-ODD TOPOLOGICAL NUMBER}

Now we turn to address the Hall conductivity in the Dirac cone regime.
According to the famous TKNN expression {\color{blue}\cite{TKNN}}, the
quantized value of the total Hall conductivity is described as 
\begin{widetext}
\begin{equation}
\sigma _{xy}=\frac{1}{\left( 2\pi \right) ^{2}i}\sum\limits_{E_{\lambda
}<E_{F}}\int\nolimits_{T_{MBZ}^{2}}\sum\limits_{\lambda }(\left\langle \frac{%
\partial u_{\lambda }}{\partial k_{x}}\right\vert \left\vert \frac{\partial
u_{\lambda }}{\partial k_{y}}\right\rangle -\left\langle \frac{\partial
u_{\lambda }}{\partial k_{y}}\right\vert \left\vert \frac{\partial
u_{\lambda }}{\partial k_{x}}\right\rangle )d\mathbf{k,}
\end{equation}
\end{widetext}where $\left\vert u_{\lambda }\right\rangle $ is the wave
function of the energy band $\lambda $ and $T_{MBZ}^{2}$ is the magnetic
Brillouin zone (MBZ). The wave function $\left\vert u_{\lambda
}\right\rangle $ forms a U$\left( 1\right) $ fibre bundle on the MBZ and the
corresponding quantized value of the Hall conductivity is the first Chern
number which is a topological invariant of the U$\left( 1\right) $ bundle.
If the Fermi energy $E_{F}$ lies in one of the energy gaps, the total
quantized value of the Hall conductivity is a sum of the first Chern numbers
corresponding to each energy band below the Fermi energy. {In Fig. 2}, using
lattice gauge theory {\color{blue}\cite{CN}, }we calculate numerically the
Hall conductivity in the relativistic regime in the condition of Fig.1 (same
indexing){. Here the laser-induced uniform abelian gauge flux plays the role
of external magnetic field, which is used to break the time-reversal
symmetry for getting the quantum Hall conductivity. In the presence of this
magnetic field, the energy spectrum of the Dirac cones are splitted into
many relativistic LLs. We also plot the relativistic LLs in each Dirac cone
and the corresponding half-odd topological number sequences in the lower
part of Fig. 2.}

In Fig. 2(a), the calculated result is for the original lattice model
without shifting the Dirac cones. The Hall conductivity associated to each
energy gap in the Dirac cone regime is $\{-10,-6,-2,2,6,10\}$, which has a
step of $4$ as that in graphene. Compared with the usual Hall conductivity,
it has an anomalous form $\sigma _{xy}=4(n+1/2)/h$ ($n=0,\pm 1,..$). This
anomaly arises from the fact that each Dirac cone contributes to the total
Hall conductivity as a half-odd topological number sequence: $n+1/2$. The
factor 4 accounts for double valley and double spin degeneracy. As indicated
in Fig. 2(a), the sum of the half-odd topological numbers over the four
Dirac cones in each gap just corresponds to the calculated Chern number.
However, because the TKNN formula guarantees that the quantized values of
the Hall conductivity for each gap are integer topological Chern numbers {%
\color{blue}\cite{TKNN}}, the contribution of a single Dirac cone to the
Hall conductivity as half-odd topological numbers has been hidden. 
\begin{figure}[tbp]
\includegraphics[width=8cm]{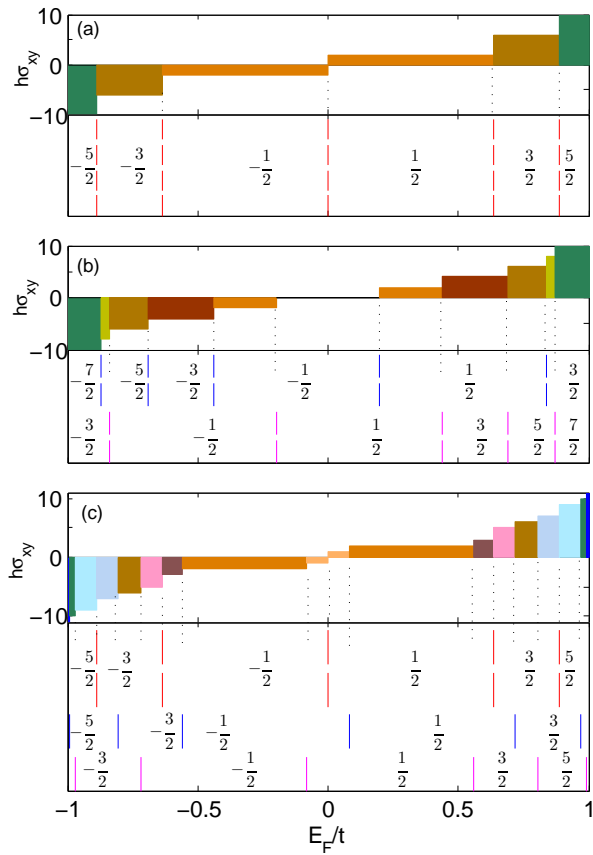}
\caption{(Color online) In the conditions of Fig.1 (same indexing), the Hall
plateaus against the Fermi energy in the relativistic regime is shown when $%
\Phi =1/121$. For comparison, in the lower part of each figure, we also plot
the relativistic LLs and show the half-odd Chern number corresponding to
each gap in the four Dirac cones respectively. Here we choose low magnetic
flux ($\Phi =1/121$) so that we can get more LLs in the Dirac cone for
rendering their contribution to the Hall conductivity more pronounced.}
\end{figure}

To manifest the contribution of the single Dirac cone, in Fig. 2(b-c), we
shift the energies of the Dirac cones and expect that the effect of this
shift on the Hall conductivity can give evidence of their own contribution.
In Fig. 2(b), the Hall conductivity is calculated when one pair of Dirac
cones is shifted equally toward positive energy and the other pair toward
negative energy, and the {relativistic LLs in each shifted Dirac cone are
also plotted in the lower part of the figure.} The calculated result shows
that the new Hall sequence is $\{-10,-8,-6,-4,-2,0,2,4,6,8,10\}$. By
comparison, one also can find that the positions of the shifted relativistic
LLs coincide with the positions of the jumps of the new Hall plateaus, i.e.
each Hall plateau corresponds to one of the new energy gaps induced by the
above shift. This means that the appearance of the new Hall sequence is just
because of the shifted Dirac cones. Furthermore, when the Fermi energy lies
in one of the new gaps, as indicated in the lower part of Fig. 2(b), each
shifted Dirac cone contributes to the total Hall conductivity of the gap
with a half-odd topological number. It is found that the sum of the four
half-odd topological numbers corresponding to each gap is equal to the
numerical calculated Chern number of the gap. For example, for the central
gap, the two positive (negative) shifted Dirac cones contribute equally to
the Hall conductivity of the gap with a half-odd topological number $-1/2$ $%
(1/2)$, their sum $2\times (-1/2+1/2)$ is just equal to the numerical
calculated Chern number $0$. As seen from this agreement, the half-odd
contribution of single Dirac cone is manifested. In Fig. 2(c), we only shift
one Dirac cone toward positive and one toward negative energy, and the other
two ones keep intact. The induced new Hall sequence $%
\{-10,-9,-7,-6,-5,-3,-2,-1,1,2,3,5,6,7,9,10\}$ becomes more complicated.
However, one can still find that the sum of the half-odd topological number
sequences in the Dirac cones with and without shift agrees with the
calculated Chern number sequence. In fact, different shifting cases make no
difference on the underlying agreement. Note that this agreement is very
surprising, because there is no obvious reason to show that the
superposition of field theory for the shifted Dirac cones can also give the
same Chern numbers {\color{blue}\cite{graphene3}}. So, based on these
agreements, we can demonstrate indirectly the contribution of single Dirac
cone as half-odd topological numbers. Due to TKNN, one can only have
indirect signatures of half-odds, but our setup allows one to have such
partial signs in a clearer and simpler fashion as compared to existing one {%
\color{blue}\cite{graphene3}}.

\begin{figure}[tbp]
\includegraphics[height=8cm,width=8cm]{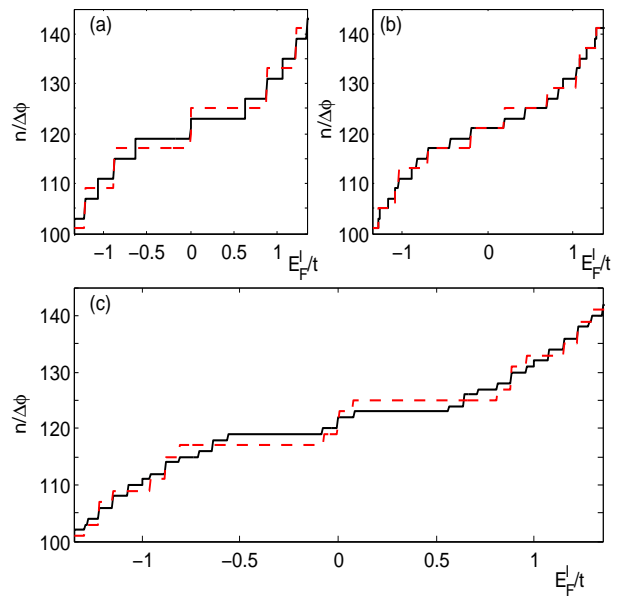}
\caption{(Color online) The atomic density profiles $n(E_{F}^{l})/\triangle
\Phi $ against the local Fermi energy for the Dirac Fermions in the
condition of Fig. 1 (same indexing) and the solid (dash) line corresponding
to $\Phi =1/121$ ($2/121$). }
\end{figure}

\section{DETECTION}

In the following, we explain how to detect the atomic Hall conductivity with
the density profile measurement {\color{blue}\cite{DP1,DP11}}. The density
profile of the trapped atoms can be measured through the time-of-flight
(TOF) imaging with the light absorption {\color{blue}\cite{DP2}} instead of
momentum distribution, which is specific to two dimensional system {%
\color{blue}\cite{DP3}}. Note that usually an external harmonic trap has
been applied to confine the atoms in the optical lattice. If the trap
potential varies slowly compared to the lattice spacing, the local density
approximation (LDA) is well satisfied {\color{blue}\cite{LDA}}. Under this
approximation, the local Fermi energy is defined as $E_{F}^{l}(r)=E_{F}-V(r)$%
, where $E_{F}$ is the Fermi energy in the trap center and $V(r)=$ $m\omega
^{2}r^{2}/2$ is the global harmonic trap potential. The atomic density at
temperature $T$ is written as

\begin{equation}
n(E_{F}^{l})=\int \frac{d\mathbf{k}}{\left( 2\pi \right) ^{2}}f(k\mathbf{,}%
E_{F}^{l})
\end{equation}%
where $f(k\mathbf{,}E_{F}^{l})=1/\{\exp [(E_{k}-E_{F}^{l})/T]+1\}$ is the
Fermi-Dirac distribution. If the local Fermi energy $E_{F}^{l}(r)$ falls in
a energy gap as plotted in Fig. 2, the atomic density $n(E_{F}^{l})$ will
depict a plateau. The Hall conductivity is related to the atomic density
according to the Streda formula $\sigma _{xy}=\frac{1}{h}\frac{\partial n}{%
\partial \Phi }|_{E_{F}^{l},T}$. As can be seen from this formula, to
measure the Hall conductivity, we need to choose two magnetic flux values,
and identify the plateaus in both density profiles that correspond to the
same gap. The density difference of the two plateaus divided by the magnetic
flux difference gives the Hall conductance of the gap. In Fig. 3, at low
temperature, we plot the density profiles of the atomic gas in the
relativistic regime in the condition of Fig. 1 (same indexing). Each case
contains two density profiles for $\Phi _{1}=1/121$ and $\Phi _{2}=2/121$
and $\triangle \Phi =\Phi _{2}-\Phi _{1}=1/121$. Through comparing the two
density profiles for $\Phi _{1}$ and $\Phi _{2}$, the atomic Hall
conductivity can be derived as $h\sigma _{xy}=(n_{2}-n_{1})/\triangle \Phi $ 
{\color{blue}\cite{DP1,DP11}}. One can find that the derived atomic Hall
conductivity from the calculated theoretical result in Fig. 3 agrees with
the prediction in Fig. 2.

\section{DISCUSSION AND SUMMARY}

Finally, we give a brief discussion on the density profile measurement from
an experimental point of view. Due to the external harmonic trap, the local
Fermi energy decreases continuously from the trap center to edge. If we let
the Fermi energy at the center of the trap higher than $t$, it automatically
scans from $t$ to $-t$. The local Fermi energy at the center of the trap can
be increased by loading more atoms in the trap {\color{blue}\cite{FH}}. It
is not difficult to make the local Fermi energy higher than $t$, as it has
been done in many previous experiment of fermions in optical lattices {%
\color{blue}\cite{FH}}. Moreover, there is no need to look for the exact
value of the local Fermi energy as long as we do the comparison between
plateau densities in the same energy gap. It means that, to a certain
degree, the measurement result is insensitive to the influence of the
experimental noise on the parameters ($\Phi ,\alpha ,\beta $), which also
reflects the topological character of the quantized value of the Hall
conductivity. Note that the measure result is also influenced by the
temperature and the external trap frequency. The temperature is related to
the visibility of plateaus in density profile. As shown in {\color{blue}\cite%
{DP1}}, the temperature of the atoms should be cooled to the order of nk for
observing discernible plateaus, which is easy than realizing the degenerate
Fermi gases. The external harmonic trap frequency determines the validity of
the LDA. If the trap potential varies smoothly so that the confinement scale
does not approach the scale of lattice constant, the LDA will be safe.
Obviously, this requirement is within the current experimental technology.

In summary, with cold atoms in a non-Abelian optical lattice, we have
proposed an experimental scheme to simulate the quantum anomalous Hall
effect as in graphene. By tuning the laser intensity to change the gauge
fluxes, we show that the contribution of a single Dirac cone to the Hall
conductivity is a half-odd topological number sequence, and such topological
numbers are detectable through the standard density profile measurement used
in cold atom systems.

\bigskip

F. Mei thanks N. Goldman, Hui Zhai and R.O. Umucal\i lar for helpful
discussions. This work was supported by the NSFC under Grant No. 60978009,
No. 10974059 and No. 11074079, the SKPBR of China (No.2007CB925204,
No.2009CB929604 and No.2011CB922104), and NUS Academic Research (Grant No.
WBS: R-710-000-008-271).

\bigskip

\bigskip

\bigskip

\bigskip

\end{document}